\documentclass[trackchanges]{aastex701}

\begin{document}


\title{{Machine Learning Confirms GW231123 is a ``Lite"-Intermediate Mass Black Hole Merger}}

\author[0000-0001-8700-3455]{Chayan Chatterjee}
\affiliation{Department of Physics and Astronomy, Vanderbilt University\\ 2201 West End Avenue, Nashville, Tennessee - 37235,}
\affiliation{Data Science Institute, Vanderbilt University\\ 1400 18th Avenue South Building, Suite 2000, Nashville, Tennessee - 37212, }
\email[show]{chayan.chatterjee@vanderbilt.edu}

\author[0009-0009-5018-848X]{Kaylah McGowan}
\affiliation{Department of Physics and Astronomy, Vanderbilt University\\ 2201 West End Avenue, Nashville, Tennessee - 37235,}
\email[show]{kaylah.b.mcgowan@vanderbilt.edu}

\author[0009-0007-9657-1412]{Suyash Deshmukh}
\affiliation{Department of Physics and Astronomy, Vanderbilt University\\ 2201 West End Avenue, Nashville, Tennessee - 37235,}
\email[show]{suyash.deshmukh@vanderbilt.edu}

\author[0009-0004-9156-8079]{Nicholas Tyler-Howard}
\affiliation{Department of Physics and Astronomy, Vanderbilt University\\ 2201 West End Avenue, Nashville, Tennessee - 37235,} 
\email[show]{nicholas-tyler.howard@vanderbilt.edu}

\author[0000-0003-1007-8912]{Karan Jani}
\affiliation{Department of Physics and Astronomy, Vanderbilt University\\ 2201 West End Avenue, Nashville, Tennessee - 37235,} 
\email[show]{karan.jani@vanderbilt.edu}

\begin{abstract}
The LIGO–Virgo–KAGRA Collaboration recently reported GW231123, a black hole merger with total mass of around $190 - 265 ~M_{\odot}$. This event adds to the growing evidence of `lite'-intermediate mass black hole (IMBH) discoveries of post-merger black holes $\gtrsim 100~M_\odot$. GW231123 posed several data analysis challenges owing to waveform‑model systematics and presence of noise artifacts called glitches. We present the first comprehensive machine learning analysis to further validate this event, strengthen its astrophysical inference, and characterize instrumental noise in its vicinity. Our approach uses a combination of tools tailored for specific analyses: \textsc{GW-Whisper}, an adaptation of OpenAI's audio transformer; \textsc{ArchGEM}, a Gaussian mixture model-based soft clustering and density approximation software; and \textsc{AWaRe}, a convolutional autoencoder. We identify the data segment containing the merger with $> 70\%$ confidence in both detectors and verify its astrophysical origin. We then characterize the scattered light glitch around the event, providing the first physically interpretable parameters for the glitch. We also reconstruct the real waveforms from the data with slightly better agreement to model‐agnostic reconstructions than to quasi-circular models, hinting at possible astrophysics beyond current waveform families (such as non-circular orbits or environmental imprints). Finally, by demonstrating high‐fidelity waveform reconstructions for simulated mergers with total masses between $100 - 1000 M_{\odot}$, we show that our method can confidently probe the IMBH regime. Our integrated framework offers a powerful complementary tool to traditional pipelines for rapid, robust analysis of massive, glitch‑contaminated events.

\end{abstract}



\section{Introduction} 

The gravitational wave (GW) event GW231123 \citep{GW231123} stands out as the most massive binary black hole (BBH) merger observed to date, with total mass in the range $ 190 - 265 M_{\odot}$ and unusually high spins ($\sim 0.9 \ \text{and} \sim 0.8 $). The event represents the latest addition to the growing catalog of low-mass or ``lite" IMBH discoveries using GWs \citep{Lite_IMBH, GW170502, GW190521_discovery}. The inferred masses of this event straddle the theorized $ 60 - 130 M_{\odot}$ pair instability mass gap, implying either a non-standard stellar origin or hierarchical mergers in dense environments, marking GW231123 as a critical probe of BH formation channels \citep{Floor&Ilya_rates}.\\

Accurate characterization of GW231123 has been challenging, mainly because significant waveform-model systematics have been observed between state-of-the-art inspiral–merger–ringdown (IMR) templates \citep{NRSur7dq4, IMRPhenomXPHM, IMRPhenomTPHM, IMRPhenomXO4a, SEOBNRv5PHM}, showing divergence in the high mass, high spin regime. In addition to that, the data around the merger was contaminated by non-Gaussian transient noise artifacts (``glitches") in both detectors. In LIGO Hanford \citep{LIGO}, a glitch related to the differential arm control loop, in a frequency range between 15 - 30 Hz had appeared at 1.1-1.7 s before the merger. This glitch was removed by using a phenomenological, wavelet-based model using the BayesWave algorithm \citep{BayesWave, BayesWave1, BayesWave2}. In LIGO Livingston \citep{LIGO}, a possible scattered light glitch \citep{glitches_definition1, glitches_definition2, glitches_definition3} was identified 2.0 - 3.0 s before the event in a frequency range between 10 - 20 Hz. Since it was determined that this glitch would have no effect on parameter inference, no subtraction was applied. \\

Recent studies have shown that broadband glitches that overlap with GW signals can induce significant biases in parameter estimation if coincident with the merger \citep{Impact_of_glitch1, Impact_of_glitch2, Impact_of_glitch3, Impact_of_glitch4, Impact_of_glitch5, Impact_of_glitch6}. Ensuring reliable data quality is paramount for an event of such astrophysical significance, particularly because certain glitches -- like Blips, Tomte, and Koi fish \citep{glitches_definition1, glitches_definition2, glitches_definition3}  -- can closely mimic the morphology of high-mass BBH signals. Traditional mitigation methods, relying on manual inspection and heuristic vetoes, are both labor-intensive and insufficiently scalable as detector data rates continue to climb. At the same time, the continual emergence of new glitch classes underscores the need for model-agnostic removal strategies \citep{LIGO_O4_performance}. In this context, machine learning (ML) techniques -- especially deep neural networks -- offer a compelling alternative, providing automated, scalable, and broadly applicable frameworks for both noise suppression and waveform reconstruction \citep{DeepExtractor, glitch1, glitch2, glitch3,glitch4, glitch5, glitch6, Chatterjee_2021, Bacon_denoising}. \\

In this work, we introduce a fully integrated ML framework tailored to high‑mass, glitch‑contaminated events like GW231123. Our pipeline is built around three complementary tools: \textsc{GW‑Whisper} \citep{chatterjee_gw_whisper_2024}, an adaptation of OpenAI’s Whisper audio transformer \citep{Whisper}, that processes whitened strain and performs low‑latency, segment‑level classification, simultaneously flagging GW signals and vetoing common glitch morphologies without human intervention. Second, we apply \textsc{AWaRe} - Attention-boosted Waveform Reconstruction network \citep{AWaRe, AWaRe_uncertainty, AWaRe_glitch}, a probabilistic convolutional autoencoder that generates uncertainty‑aware waveform reconstructions of GW signals in the time-domain.  We assess its robustness through extensive Monte‑Carlo injections that embed GW231123‑like signals in a range of glitch environments and extend to binary mergers with total masses from $100$ to $1000\,M_\odot$. Finally, \textsc{ArchGEM} \citep{McGowan2025archgem} employs Gaussian‑mixture clustering on Q‑transform spectrograms to isolate scattered‑light arches and derive physically meaningful parameters -- such as recurrence frequency, displacement, and velocity -- providing actionable feedback for detector commissioning.  Together, these components form a rapid and scalable pipeline that complements traditional matched filter \citep{Matched_filtering} and Bayesian analyses, enabling more reliable signal validation and noise characterization for current and future observations of IMBH mergers. \\

This paper is organized as follows: Section 2 describes results obtained using \textsc{GW-Whisper} on 8 seconds of whitened data around the event. Section 3 explores scattered light glitch characterization around GW231123 using \textsc{ArchGEM} with comparisons against trends observed across the O4 observation run. Section 4 presents analyses performed using \textsc{AWaRe} on GW231123 data and simulated injection sets that span a wide range of IMBH masses. In Section 5, we present validation and robustness results across all three models using a simulated injection set. We present our conclusions in Section 6. \\


\section{Signal Classification}

\begin{figure*}[t!]
    \centering
    \includegraphics[width=1\textwidth]{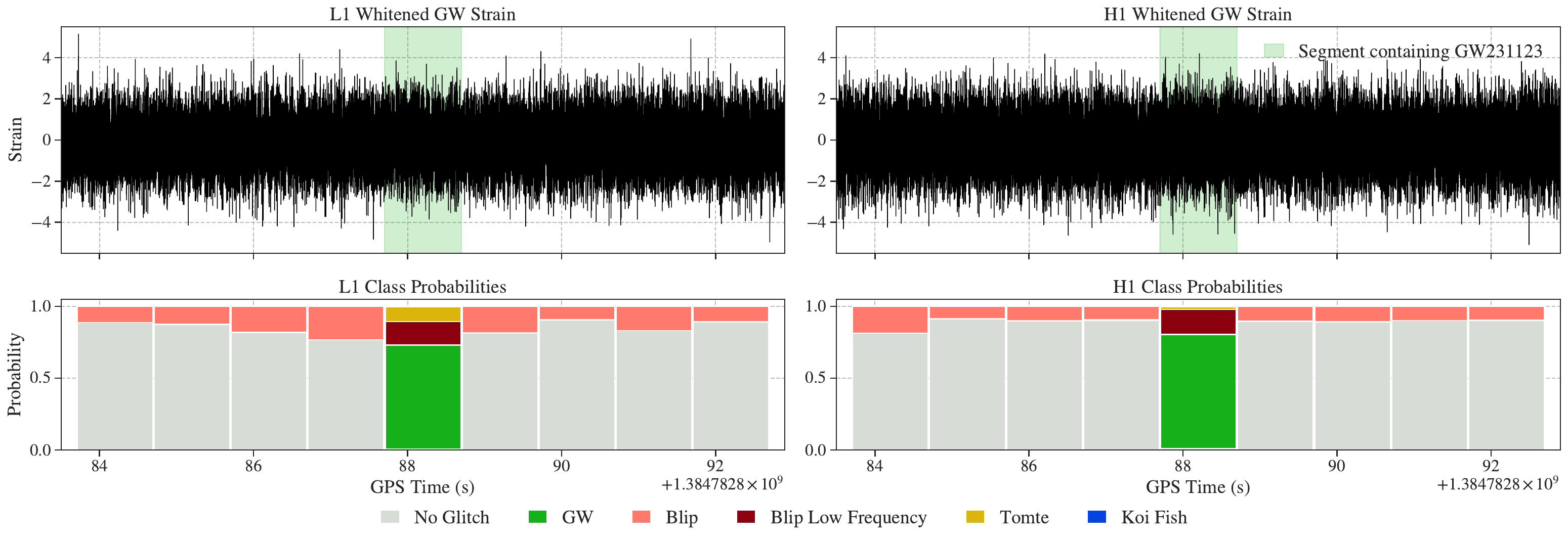}
    \caption{Predicted class label probabilities from 8 seconds around GW231123 using \textsc{GW-Whisper} in Livingston (left) and Hanford (right). The top row shows the whitened strains from the respective detectors with the segment containing the event highlighted green. The bottom row contains the probabilities for each label in each segment. The model predicts ``No Glitch" for all segments around the event and ``GW" in the segment that contains the event.}
    \label{fig:whisper}
\end{figure*}

We use \textsc{GW-Whisper} \citep{chatterjee_gw_whisper_2024} to classify and tag glitches and GW events present in 8 s of data around the merger time of GW231123. \textsc{GW-Whisper} builds on OpenAI’s Whisper audio transformer architecture \citep{Whisper}, originally trained on 680,000 hours of audio data. Chatterjee et al. \citep{chatterjee_gw_whisper_2024} demonstrated that it is possible to adapt Whisper -- using the \textit{Whisper-tiny} model with 39 million parameters -- for GW data analysis by fine-tuning only $\sim$0.5\% of the parameters of the pre-trained Whisper encoder on GW injections and glitch data. This adaptation was achieved using a parameter-efficient fine-tuning \citep{PEFT} technique called DoRA (Weight-Decomposed Low-Rank Adaptation) \citep{DoRA}, in which the original transformer weight matrices of dimension $(d \times d)$ are frozen, and additional low-rank matrices of dimension $(d \times r)$ and $(r \times d)$—where $r$ is the decomposition rank—are trained during fine-tuning. In this work, we used a rank of $r = 8$. Two versions of \textsc{GW-Whisper} were introduced by \citet{chatterjee_gw_whisper_2024}: one for signal detection and one for glitch classification. Here, we report results from the glitch-classification model. In order to perform GW and glitch classification, \textsc{GW-Whisper} was trained on a curated set of glitches from the GravitySpy catalog \citep{GravitySpy} with similar morphology to high-mass GW events: Blip, Low-Frequency Blip, Koi Fish, and Tomte, and a simulated set of BBH mergers with total mass $> 50 M_\odot$ injected into background O3 data. An additional ``No Glitch” class was included to represent background segments without prominent glitches or GW signals (see Fig.~4(c) in \cite{chatterjee_gw_whisper_2024} for reference). \\

We employ this trained model to identify GW231123 and surrounding glitches. The model takes as input log-mel spectrogram representations of single-detector, whitened, 1 s strain segments and outputs a probability distribution over the aforementioned set of GW+glitch classes for each segment. Fig.~\ref{fig:whisper} shows the model outputs for 8 s data surrounding GW231123. In Hanford and Livingston data, \textsc{GW-Whisper} identifies the segment containing the signal with a confidence of 79.32\% and 72.33\% respectively. All other segments are consistently classified as ``No Glitch” with high confidence (there is a small residual probability in these segments and we find that they are predominantly assigned to the ``Blip” class). Although both detectors contained low-frequency glitches near the time of the event, these transients were removed by the 20 Hz high-pass filter applied in the data pre-processing step. This filtering step matches the processing used for generating the training dataset, and is intended to remove the low-frequency noise component in the data, enabling the model to better focus on GW events. Training the glitch classification model takes around 5 hrs on a single NVIDIA DGX 80 GB A100 GPU. Inference using the trained model takes a few milliseconds on the same device. \\

\begin{figure*}[b!]
    \centering \includegraphics[width=1.1
    \textwidth]{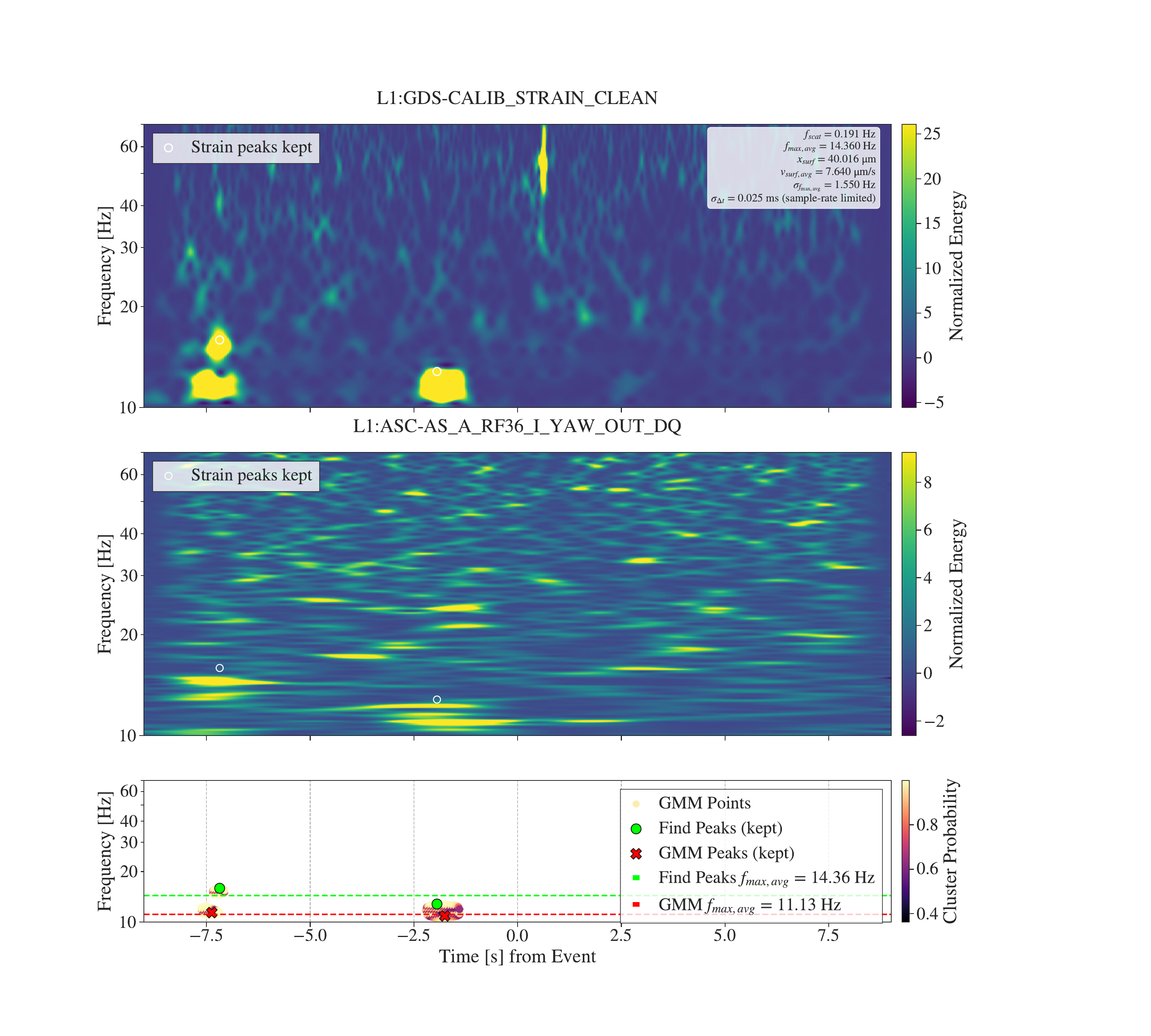}
\caption{The top panel displays the Q-transform visualization of the GW231123 event and scattered light glitches observed at the Livingston Observatory. The middle panel displays the scattered light glitches propagating in an Alignment Sensing and Control (ASC) auxiliary channel. The bottom panel is the \textsc{ArchGEM} output highlighting the software's capability to recognize scattered light morphology and extract them for further analysis using a dual methodology ML approach.}
\label{fig:archout}
\end{figure*}

\section{Scattered Light Glitch Characterization}
\label{sec:scatteredlight}

To investigate low-frequency scattering noise in aLIGO preceding the GW23123 event, we applied the \texttt{\textsc{ArchGEM}} software to strain data from the Livingston interferometer, using the calibrated channel \texttt{L1:GDS-CALIB\_STRAIN\_CLEAN}. Our analysis focused on a 16-second window centered on GPS time 1384782888. The upper panel of Fig.~\ref{fig:archout} presents the resulting Q-transform spectrogram, which displays a series of arch-like features below 20~Hz, typical of scattered-light artifacts caused by moving surfaces coupling into the main interferometric beam path. We used a two-pronged analysis to isolate and characterize these arches. First, we employed a peak-finding algorithm to extract the local maxima in frequency over time for high-energy pixels. Second, we applied Gaussian Mixture Modeling (GMM) with 9 components to cluster time-frequency-energy points and identify statistically distinct groupings of noise activity. Both methods included a post-processing filter to retain only the most temporally distinct peaks. The lower panel of Fig.~\ref{fig:archout} overlays the retained peaks from both methods. The ``Find Peaks'' method identified several high-confidence features, yielding an average maximum arch frequency of $f_{\mathrm{max,avg}} = 14.36$~Hz. The GMM method identified centroids clustered around lower frequencies, with a corresponding average of $f_{\mathrm{max,avg}} = 11.13$~Hz. \\

From the distribution of the retained peak times, we infer a scattering recurrence frequency of $f_{\mathrm{scat}} = 0.19$~Hz, indicating a periodic process consistent with a surface oscillating at low frequency. Using the measured $f_{\mathrm{scat}}$, we calculate a scattering surface displacement of $x_{\mathrm{surf}} = 40.01~\mu\mathrm{m}$ and an average surface velocity of $v_{\mathrm{surf,avg}} = 7.64~\mu\mathrm{m/s}$. We also report a standard deviation in peak frequency of $\sigma_f = 1.55$~Hz. The inter-peak time intervals exhibit a timing spread of $\sigma_{\Delta t} = 0.025$~s, which reflects the finite sampling precision of the data ($f_s = 16384$~Hz) rather than true variability in the period. Because only two cycles are observed, the period is defined by those two maxima; thus, there is effectively no statistical uncertainty in $\Delta t$, only timing resolution uncertainty from the sample rate. \\ 

It should be noted that the inferred parameters $x_{\mathrm{surf}}$ and $v_{\mathrm{surf,avg}}$ are derived under this limited-cycle assumption, and therefore represent point estimates rather than confidence intervals. Future analyses should incorporate uncertainty propagation and test for consistency of these features across auxiliary channels (e.g., seismometers, angular sensors) to strengthen the identification of scattered-light coupling. In this study, we compared the strain spectrogram to the auxiliary channel \texttt{L1:ASC-AS\_A\_RF36\_I\_YAW\_OUT\_DQ}, a known witness of angular scattering, which exhibited coincident arch-like structures at similar times and frequencies. This correspondence supports the interpretation of the event as scattered light, though additional phase-coherence testing will be required for definitive confirmation. \\

Nonetheless, the $f_{\mathrm{max,avg}}$ measured for this scattering event is lower than the mean distribution seen in ArchGEM’s O4 scattered-light study \citep{McGowan2025archgem}. The propagation of similar arches in auxiliary channels associated with scattering, as shown in Fig.~\ref{fig:archout}, supports this interpretation. ArchGEM also demonstrates strong reliability in identifying the true period of scattered-light arches while successfully ignoring the coincident GW signal itself. These results demonstrate \textsc{ArchGEM}’s ability to extract physically meaningful scattering parameters from non-linear noise features in GW strain data. The combination of clustering and peak-tracking approaches enhances robustness and interpretability, particularly for characterizing complex noise morphologies near GW event times. \\


\begin{figure*}[ht!]
\gridline{\fig{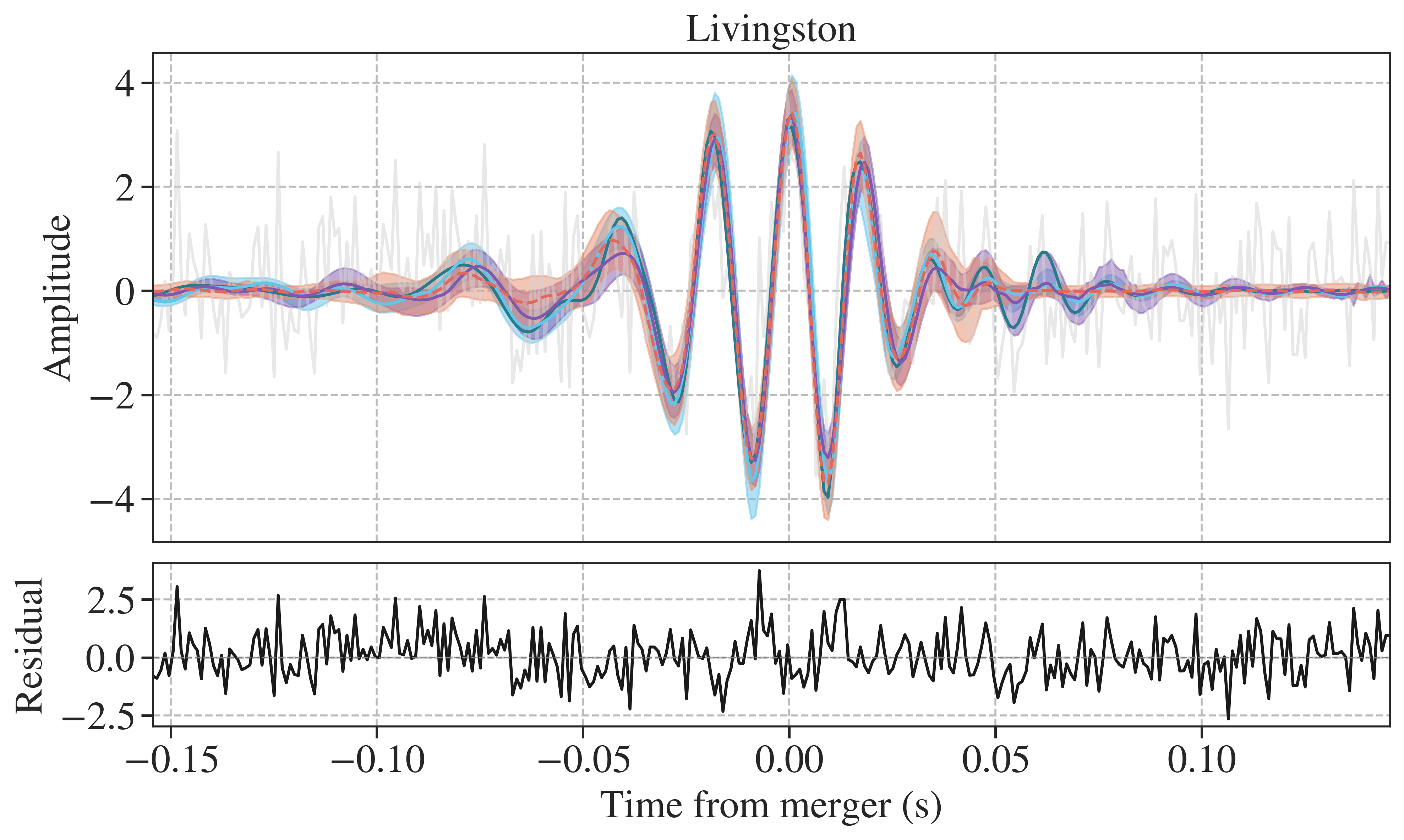}{0.5\textwidth}{(a)}
\fig{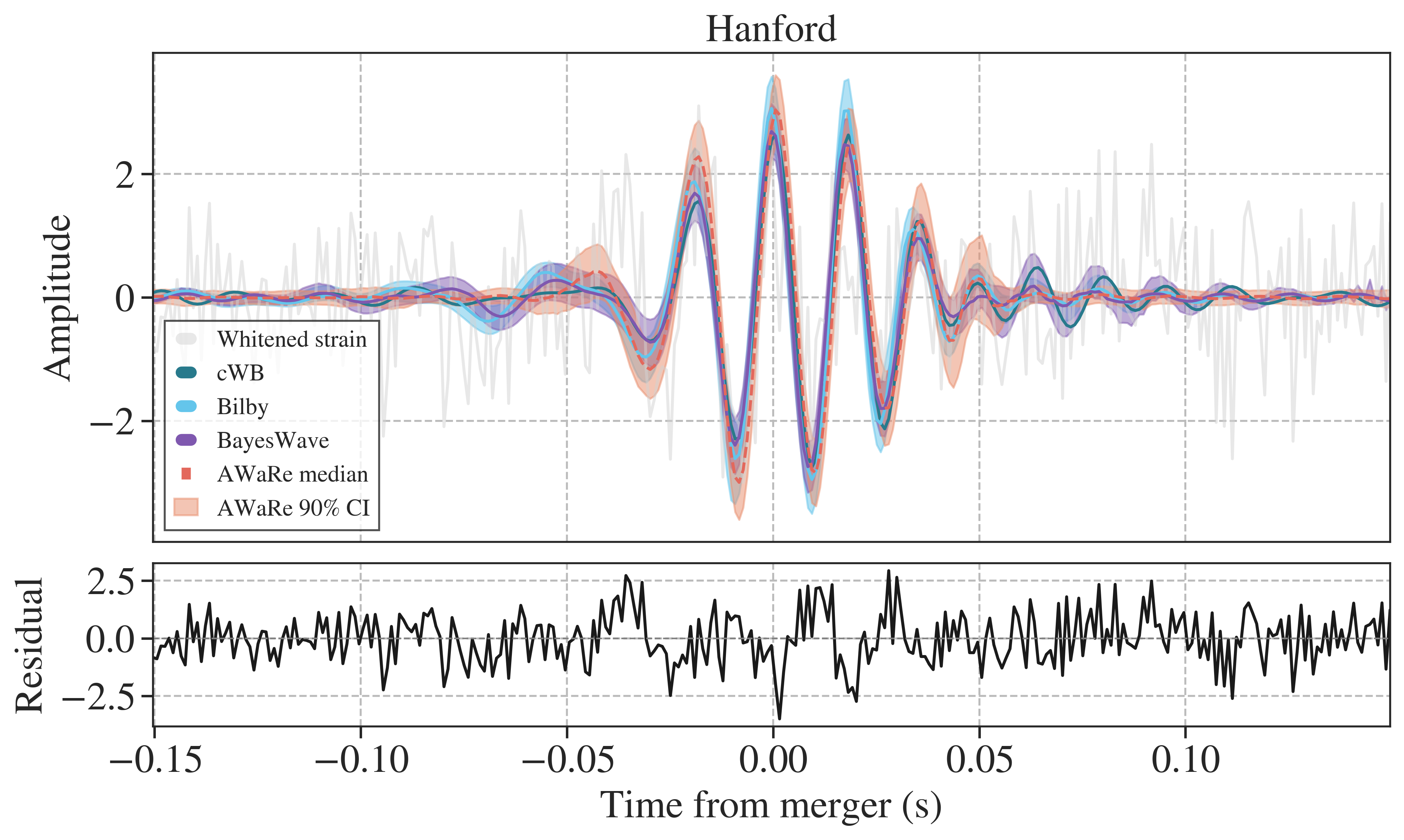}{0.5
  \textwidth}{(b)}
}
\caption{\label{fig:AWaRe_GW231123} Top: Reconstructions of (a) Livingston and (b) Hanford data of GW231123 using \textsc{AWaRe}. The red dashed curves shows the \textsc{AWaRe} mean reconstructions. The red bands show the associated reconstruction uncertainties. Reconstructions from cWB, Bilby and BayesWave are shown in green, blue and purple respectively. The whitened noisy strains are shown in grey. Bottom: Residuals obtained by subtracting the \textsc{AWaRe} mean reconstructions from the whitened strains.}

\end{figure*}

\section{Waveform reconstruction}


In previous studies, the feasibility of \textsc{AWaRe} was demonstrated for waveform reconstructions of real GW events \citep{AWaRe}. To obtain the results in this study, we trained a new version of \textsc{AWaRe} on simulated BBHs with total masses between 100 and 1000 $M_{\odot}$ in the detector frame. These signals were injected into random segments of background data around GW231123, excluding the event time. In our new model architecture, we replace the simple convolutional encoder and Long Short Term Memory (LSTM) \citep{LSTM} decoder in the original model architecture with purely convolutional U-Net \citep{UNet} encoder and decoder. The encoder U-Net consists of three levels of 1D convolutional blocks \citep{CNN_1, CNN_2}, with a multi-head self-attention module \citep{Transformer} at the narrowest ``bottleneck” layer, that produces a low-dimensional representation of the input. The signal is then rebuilt to its original length (with noise subtracted) using learned up‑scaling convolutional layers in the decoder U-Net. Instead of outputting a single waveform, the network predicts, for each time sample, the mean and spread of a Gaussian. The final reconstruction is thus a sum of independent Gaussian random variables, providing a best‑estimate waveform with its associated reconstruction uncertainty. In our setup, we used three encoder–decoder levels with 64 base filters, small 1-D kernels (size 7), and a 4-head attention block at the bottleneck. The model is trained with AdamW optimizer (learning rate $10^{-4}$, batch size 256) for 250 epochs on 1-s windows of whitened strain (sample rate 1024 Hz; 20 Hz high-pass, IMRPhenomXPHM \citep{IMRPhenomXPHM} waveform approximant). The model was trained on 250,000 samples (injections + pure noise), and 50,000 was used as validation set. The total training time was around 4 hrs on a NVIDIA DGX 80 GB A100 GPU. \\

The red dashed curves in Fig.~\ref{fig:AWaRe_GW231123} (a) and (b) show the mean \textsc{AWaRe} reconstructions, and the variances of the Gaussians are used to obtain the reconstruction uncertainty (red bands). For comparisons, we plot the waveform-model agnostic cWB reconstructions (green) \citep{cWB}, reconstructions from NRSur7dq4 waveform-model parameter estimation results using the software package, Bilby \citep{Bilby} (blue) and wavelet-based reconstruction algorithm BayesWave \citep{BayesWave, BayesWave1, BayesWave2} (purple). We observe excellent amplitude and phase consistency between \textsc{AWaRe} reconstructions and the other approaches. For the Hanford (Livingston) signal, the mean \textsc{AWaRe} reconstruction shows overlaps of 92\% (97\%), 91\% (97\%) and 96\% (98\%) with cWB \citep{cWB}, Bilby \citep{Bilby} and BayesWave \citep{BayesWave, BayesWave1, BayesWave2} reconstructions respectively, indicating high accuracy. The slightly stronger agreement with the template‑free cWB and BayesWave methods -- compared with the template‑based Bilby result -- suggests that \textsc{AWaRe} captures intrinsic signal morphology that is robust to waveform‑model systematics. \\

In the bottom panel of Fig.~\ref{fig:AWaRe_GW231123}, we plot the residuals obtained by subtracting the \textsc{AWaRe} mean reconstructions from the noisy strain data. For perfect reconstruction, the residuals would be normally distributed. After obtaining the residuals, we inspect their normality by performing the Shapiro-Wilk test \citep{Shapiro_Wilk_test} on the residuals of Hanford and Livingston data and computing its p-value. Lower p-values indicate a lower probability that the coherent power in the residuals is due to instrumental noise alone. We obtain p-values of 0.671 for Hanford and 0.454 for Livingston residuals. Since both values are $ > $ 0.05, the test indicates strong evidence that the residuals follow the expected normal distribution. The optimal SNRs of the residuals in Hanford and Livingston were found to be 0.82 and 0.55, further supporting our findings that subtracting the reconstruction from the original strain leaves negligible residual power.  \\



\begin{figure}
\gridline{
\fig{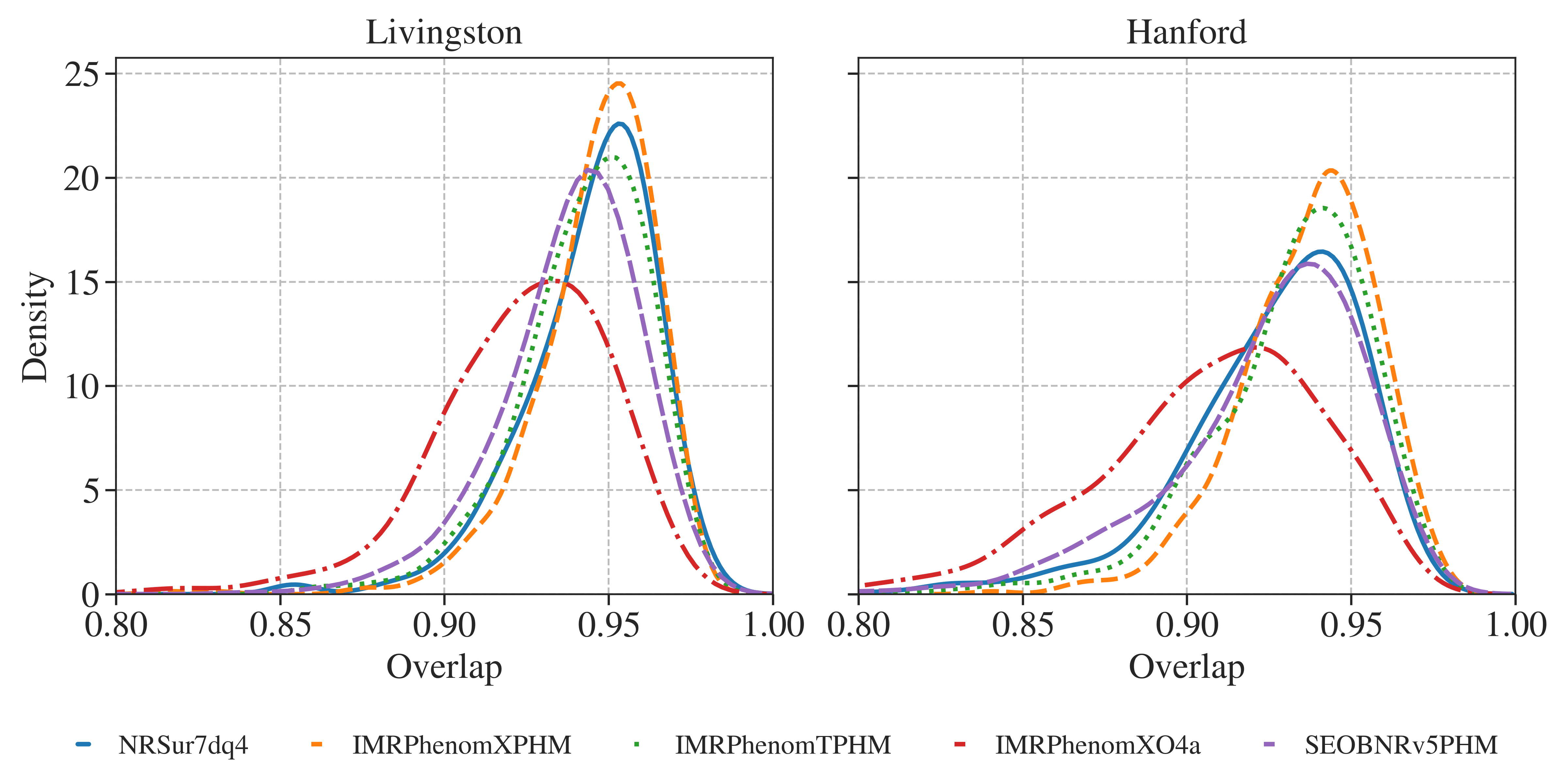}{0.8
\textwidth}{(a)}}
\gridline{
\fig{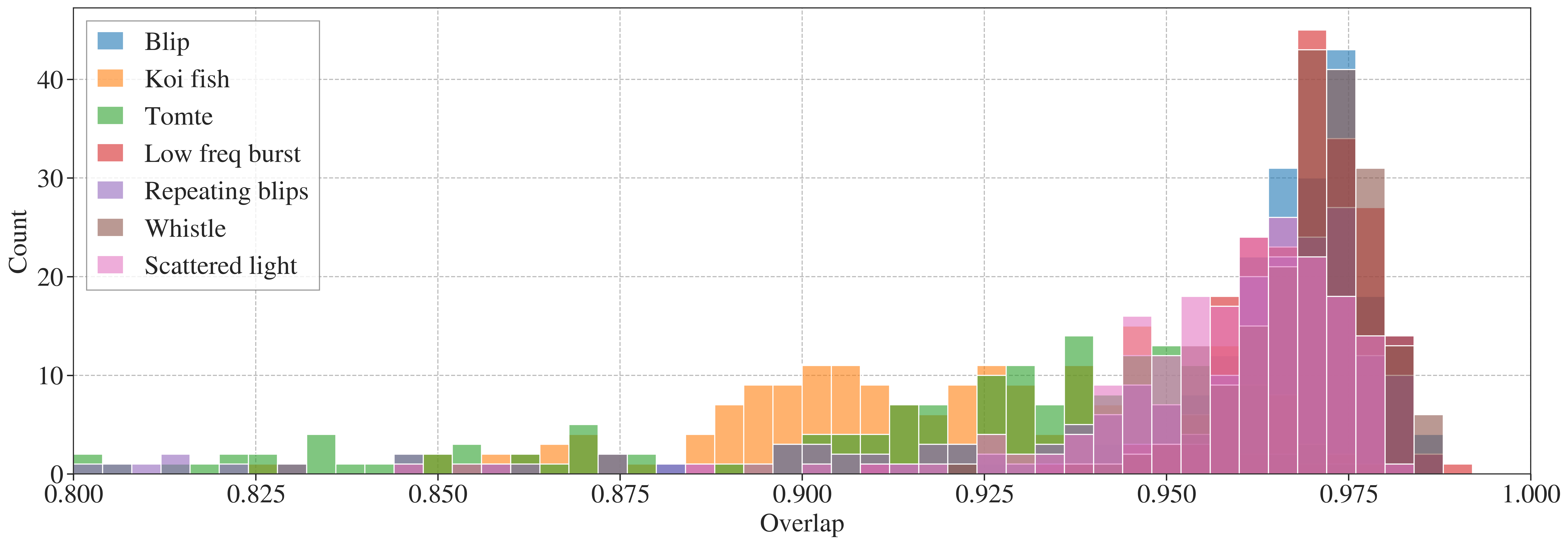}{0.8\textwidth}{(b)}}


\caption{\label{fig:AWaRe_injections} (a) Overlap distributions between GW231123‑like waveforms in Hanford and Livingston generated from posterior samples of various approximants and their \textsc{AWaRe} reconstructions.
(b) Overlap distributions between \textsc{AWaRe} reconstructions and GW231123-like injections in O3 noise contaminated by different glitches.}
\end{figure}

Next, we performed tests on simulated GW231123-like waveforms injected in noise background around the event. These waveforms were generated using Bilby posterior samples obtained using the following waveform approximants: NRSur7dq4 \citep{NRSur7dq4}, IMRPhenomXPHM \citep{IMRPhenomXPHM}, IMRPhenomTPHM \citep{IMRPhenomTPHM}, IMRPhenomXO4a \citep{IMRPhenomXO4a}, SEOBNRv5PHM \citep{SEOBNRv5PHM}. The network SNR of these injections were fixed at the reported value, i.e., 22.6. In Fig.~\ref{fig:AWaRe_injections} (a), we plot the overlap distributions between 500 simulated GW231123-like waveforms and their \textsc{AWaRe} reconstructions in Hanford and Livingston data. For all the models, the distributions peak at overlap values $>$ 0.90, indicating high reconstruction consistency between injected and recovered waveforms. \\


We also tested the performance for signals randomly injected in 1 s O3 data segments containing glitches. The motivation of this study was to test the robustness of \textsc{AWaRe} reconstructions in the presence of morphologically-similar glitches. We injected synthetic GW231123 signals from the NRSur7dq4 model into O3 strain segments containing seven common glitch morphologies (Blip, Koi-fish, Tomte, Low-frequency burst, Repeating blips, Whistle, and Scattered light). Since \textsc{AWaRe} is trained to isolate and reconstruct only the astrophysical waveform, the model yields overlap distributions sharply peaked near 1.0 for all glitch types, demonstrating its capacity to remove diverse non-Gaussian artifacts and recover the true signal with high fidelity (Fig.~\ref{fig:AWaRe_injections} (b)). Small broadening of the peaks for Tomte and Koi-fish glitches reflects slightly greater reconstruction uncertainty when these morphologies overlap the signal’s time–frequency support. These results further demonstrate how \textsc{AWaRe} is complementary to GW‑Whisper and ArchGEM: while GW‑Whisper excels at identifying the presence of a signal and classifying common glitch types, and ArchGEM specializes in characterizing scattered‑light artifacts, \textsc{AWaRe} is trained to isolate and reconstruct only the astrophysical waveform, allowing it to recover true GW signals even in the presence of any glitch morphology.\\

\begin{figure*}[b!]
    \centering \includegraphics[width=0.7\textwidth]{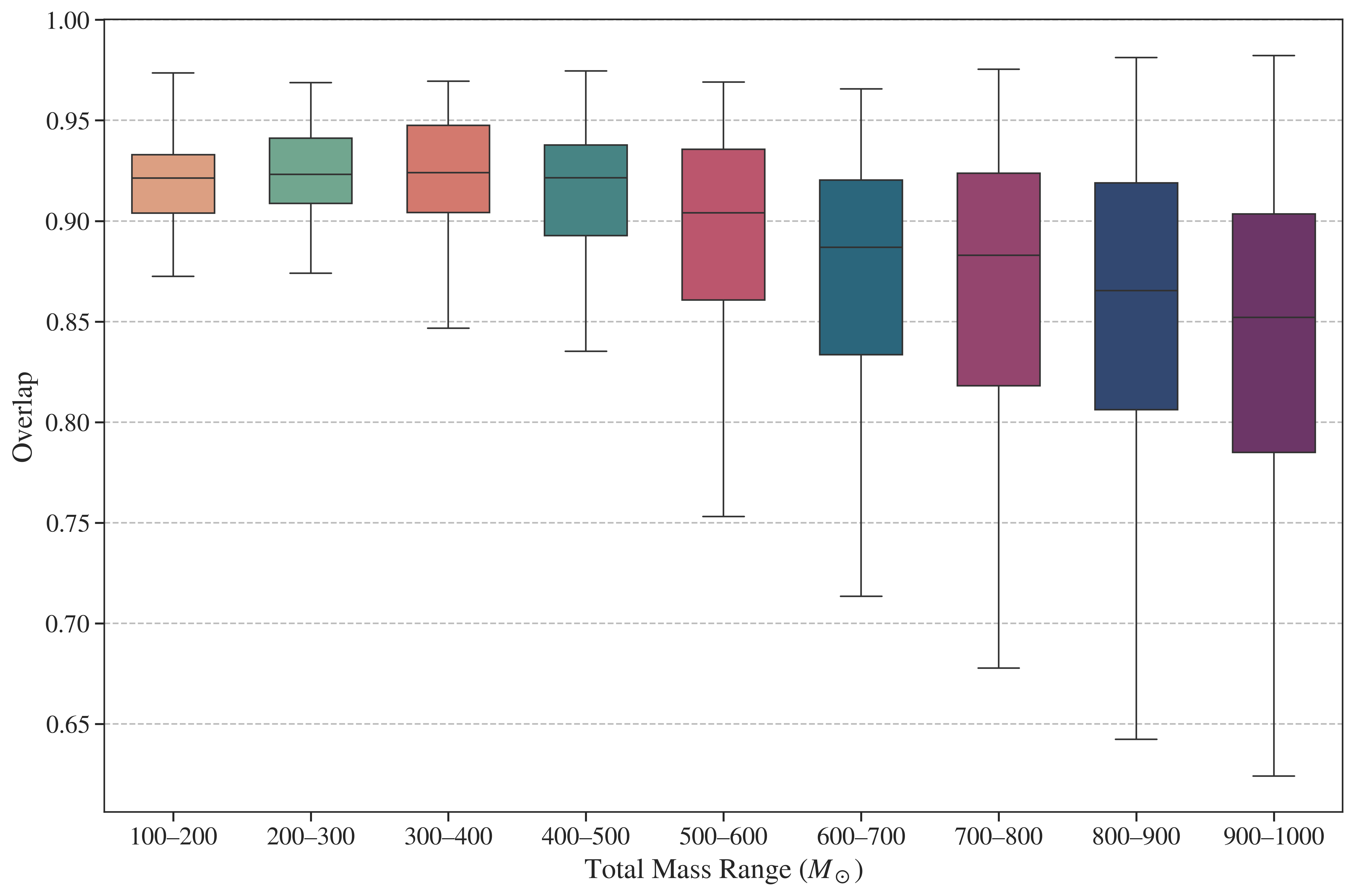}
\caption{Box plots showing the median and interquartile ranges of the overlap distributions between \textsc{AWaRe} reconstructions and original waveforms for simulated BBH events with total masses between $100 - 1000 M_{\odot}$.}
\label{fig:IMBH_test}
\end{figure*}

To test whether the \textsc{AWaRe} reconstruction pipeline generalizes beyond a single extreme event, we generated an injection set of BBH signals with total masses $100-1000 M_{\odot}$ and SNR=15, and obtained their reconstructions using \textsc{AWaRe}. The chosen mass range spans the crucial pair‑instability gap and the lower IMBH population, where signals are short, templates disagree, and glitches can masquerade as mergers. Fig.~\ref{fig:IMBH_test} shows box plots of overlap between reconstructions and true waveforms in $100 M_{\odot}$ bins. We find that the  medians remain high ($>$ 0.9) up to $~500 M_{\odot}$, then gradually fall toward $\sim 0.85 $ with broader interquartile ranges at higher masses, reflecting the increasing difficulty of recovering merger–ringdown–dominated signals in noisy data. Even in the $900-1000 M_{\odot}$ bin, most overlaps exceed 0.8, indicating substantial fidelity. These results demonstrate that \textsc{AWaRe} can rapidly and reliably vet candidate IMBH mergers and quantifies where waveform systematics and non-Gaussian noise begin to erode accuracy. \\

\section{Validation and Robustness Analysis}

To assess the reliability and generalizability of our integrated machine-learning pipeline, we conducted an extensive validation campaign using a set of simulated injections. This study spans the full analysis chain -- GW-Whisper detection, ArchGEM scattered-light characterization, and AWaRe waveform reconstruction -- under realistic detector conditions. The injections cover a broad parameter space of total masses (50–1000 M$_{\odot}$) and network SNRs (8–40), and include features such as precession, higher-order modes and eccentricity, simulated using the IMRPhenomXPHM and NRSur7dq4 and EccentricTD \citep{EccentricTD} waveform families. For eccentric binaries, we included waveforms with non-zero initial eccentricities (0.0–0.5) at 20 Hz. We also included signals contaminated with representative glitches drawn from Gravity Spy catalog. The background noise samples and glitches for this analysis were drawn from public O3 data. By comparing reconstruction overlaps, empirical coverage of predicted uncertainties, and background false-alarm rates, we quantify the robustness, calibration, and statistical performance of all three models across diverse astrophysical and noise scenarios. \\

\begin{figure*}[ht!]
\gridline{\fig{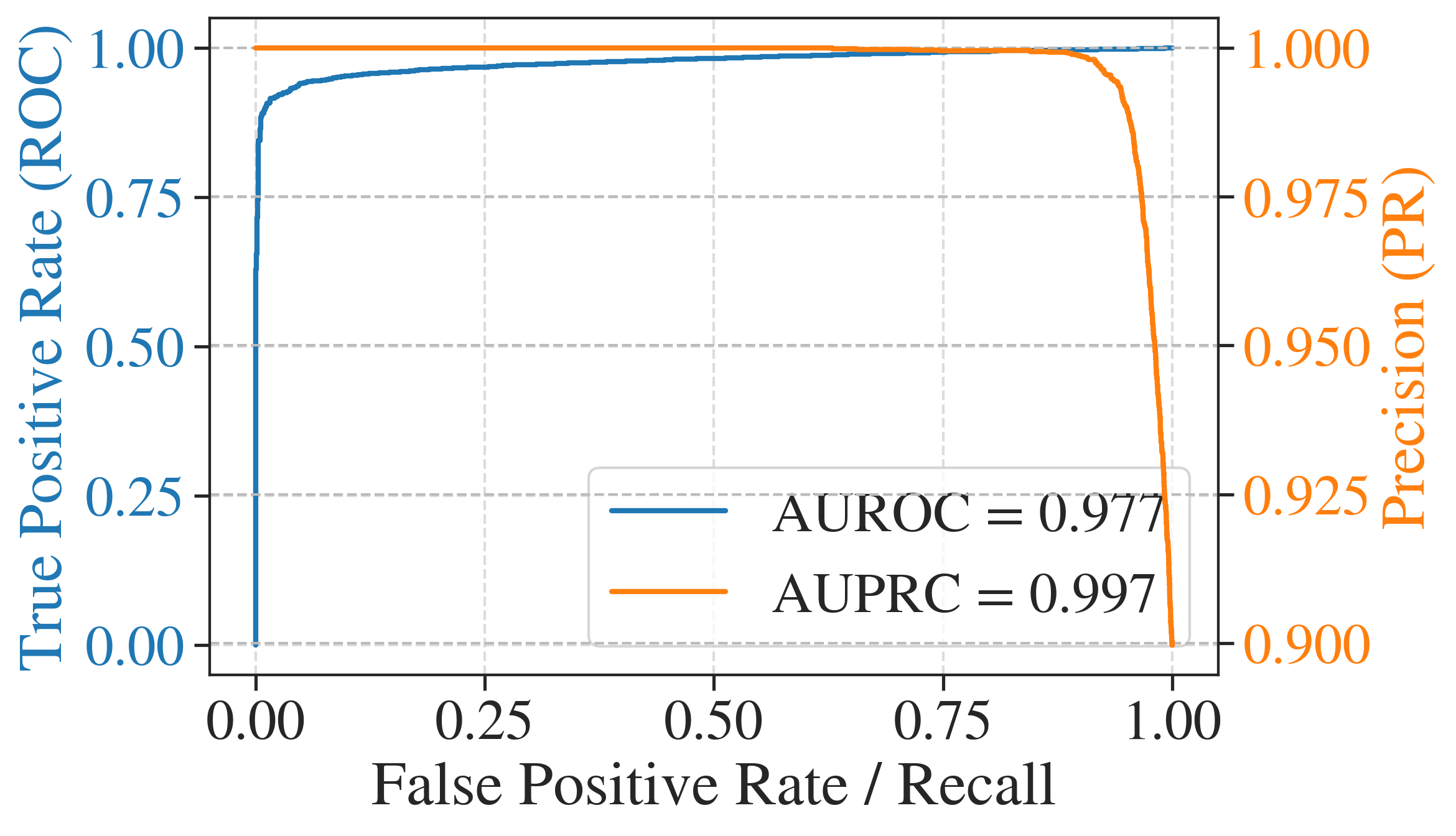}{0.52\textwidth}{(a)}
\fig{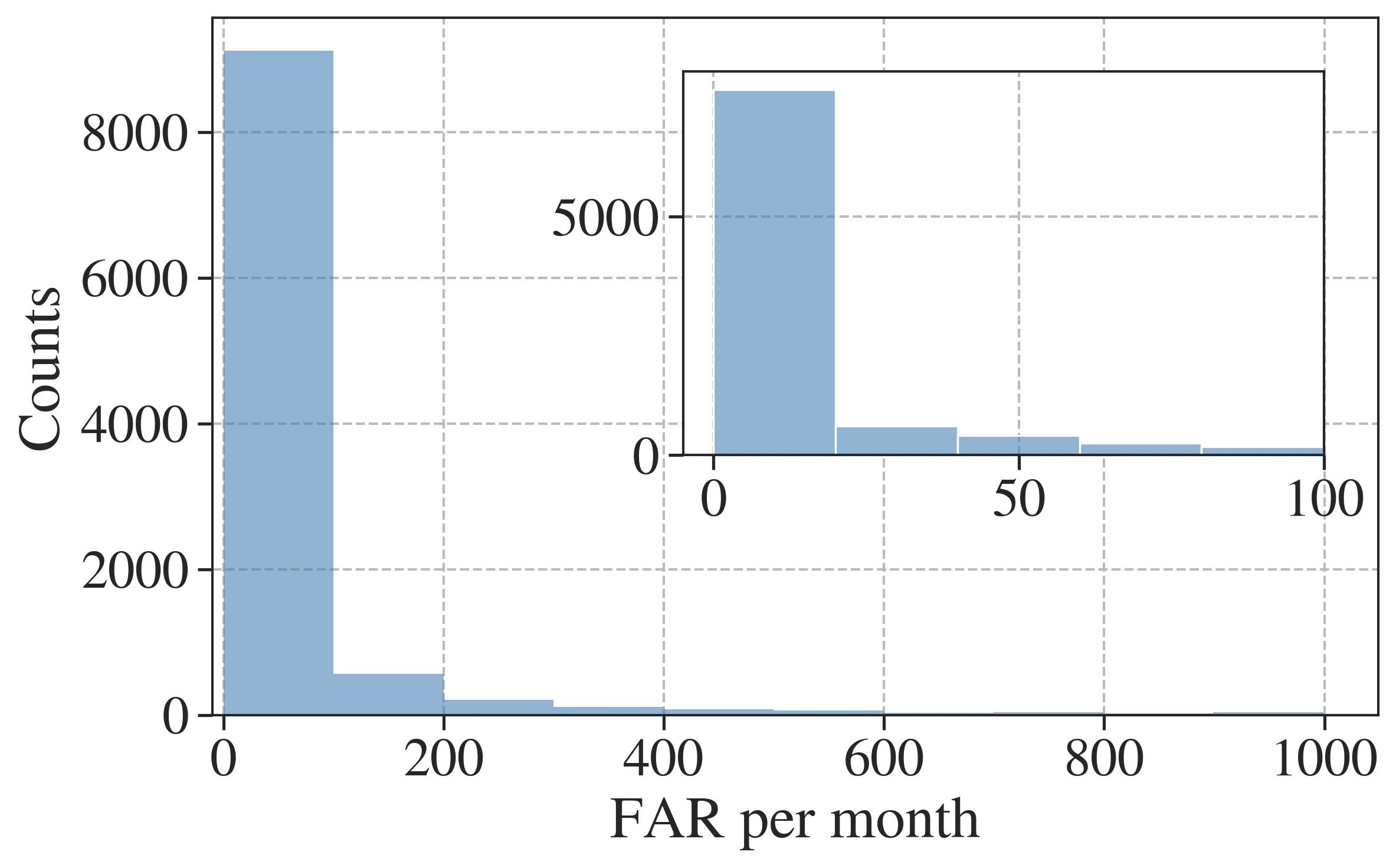}{0.48
  \textwidth}{(b)}
}
\caption{\label{fig:GW-Whisper_detection} (a) Receiver operating characteristic (ROC) and precision–recall (PR) curves obtained on the IMBH test set, after fine-tuning the original \textsc{GW-Whisper} model, initially trained for evaluation the MLGWSC-1 dataset. (b) Distribution of false-alarm rates (FAR) per month, derived from background triggers obtained by applying the fine-tuned \textsc{GW-Whisper} to dataset 4 of MLGWSC-1.}
\end{figure*}

Fig.~\ref{fig:GW-Whisper_detection}(a) shows the receiver operating characteristic (ROC) and precision–recall (PR) curves obtained after fine-tuning the original \textsc{GW-Whisper} model -- initially trained on generic BBH datasets from the first Machine Learning Gravitational-Wave Search Mock Data Challenge (MLGWSC-1) \citep{ML_MDC} -- using 200,000 simulated injections with total masses between 100 and 1000~M$_{\odot}$. The fine-tuned model achieves an area under the ROC curve (AUROC) of 0.977 and an area under the precision–recall curve (AUPRC) of 0.997, demonstrating excellent classification performance. This experiment was designed to test the generalizability of \textsc{GW-Whisper}: despite being originally optimized for generic BBH mergers, the architecture performs robustly on the more massive IMBH regime. These results highlight the strong transferability of the model’s learned representations. A dedicated IMBH search using a modified version of \textsc{GW-Whisper}, specifically optimized for the IMBH parameter space, will be presented in a forthcoming paper. \\

To estimate the false-alarm rate (FAR) for our injection campaign, we followed the same background-estimation procedure described in \citep{chatterjee_gw_whisper_2024} and the MLGWSC-1 analysis. We ran the fine-tuned \textsc{GW-Whisper} model on the 1-month pure background noise of dataset 4 of MLGWSC-1 and obtained the model outputs. These outputs were post-processed using a clustering algorithm that groups temporally adjacent triggers (within $\Delta t = 0.2$ s) into coherent events, with the maximum detection statistic within each cluster taken as its ranking statistic. The FAR corresponding to a given detection threshold $R$ is then obtained by counting the number of background events with statistic $\geq R$ and dividing by the total background live time (one month). For events below FAR of 1 per month, we fitted an exponential tail to the upper 5\% of the background distribution to obtain continuous extrapolated FAR values. The resulting distribution of FARs per month is shown in Fig.~\ref{fig:GW-Whisper_detection} (b). About 45\% of the injections are recovered at FAR $<$ 1 per month, with a rapidly decaying tail toward higher values. The inset highlights that more than 95\% of events below fall within $\mathrm{FAR}<50$ per month, confirming that the model maintains a low false-alarm background even when tested on signals beyond its original training mass range. These results further support the robustness and generalizability of \textsc{GW-Whisper} for IMBH-scale signals. Fine-tuning the model on 200,000 samples (injections + pure noise) took around 3.5 hrs on a single DGX A100 80 GB GPU. The background estimation step on the 1 month MLGWSC-1 dataset took around 18 hrs on the same architecture. \\

To evaluate the accuracy and robustness of \textsc{ArchGEM}, we conducted a controlled simulation study using synthetic scattered-light injections designed to reproduce the dominant features of real glitches observed in GW strains. For this analysis, 9-s strain data was generated, as opposed to 1-s inputs considered for GW-Whisper and AWaRe. This choice was made to ensure the characteristics of the periodic scattering arches are appropriately captured by the algorithm. The injections for this study were generated using a dedicated simulation pipeline, which combines sinusoidal amplitude and phase modulations of the interferometer strain with tunable frequencies, amplitudes, and durations, implemented by the \textsc{GlitchPop} library \citep{GlitchPop}, and based on the scattering glitch models from \citep{Udall_scattering}. Each synthetic strain reproduces the morphology of scattered-light frequency structures in Q-transform spectrograms. Randomized phase offsets and O4a-like background noise realizations were added to emulate realistic detector conditions, ensuring that the resulting data closely resemble the mixed spectral environment of an actual observing run. The subsequent analysis is driven primarily by ArchGEM’s Gaussian Mixture Model (GMM) method, which identifies and classifies energy clusters in the Q-transform. Only peaks retained by the GMM selection are used in computing the final scattering metrics, ensuring statistical robustness and minimizing false positives. \\

\begin{figure*}[ht!]
    \centering \includegraphics[width=1.0\textwidth]{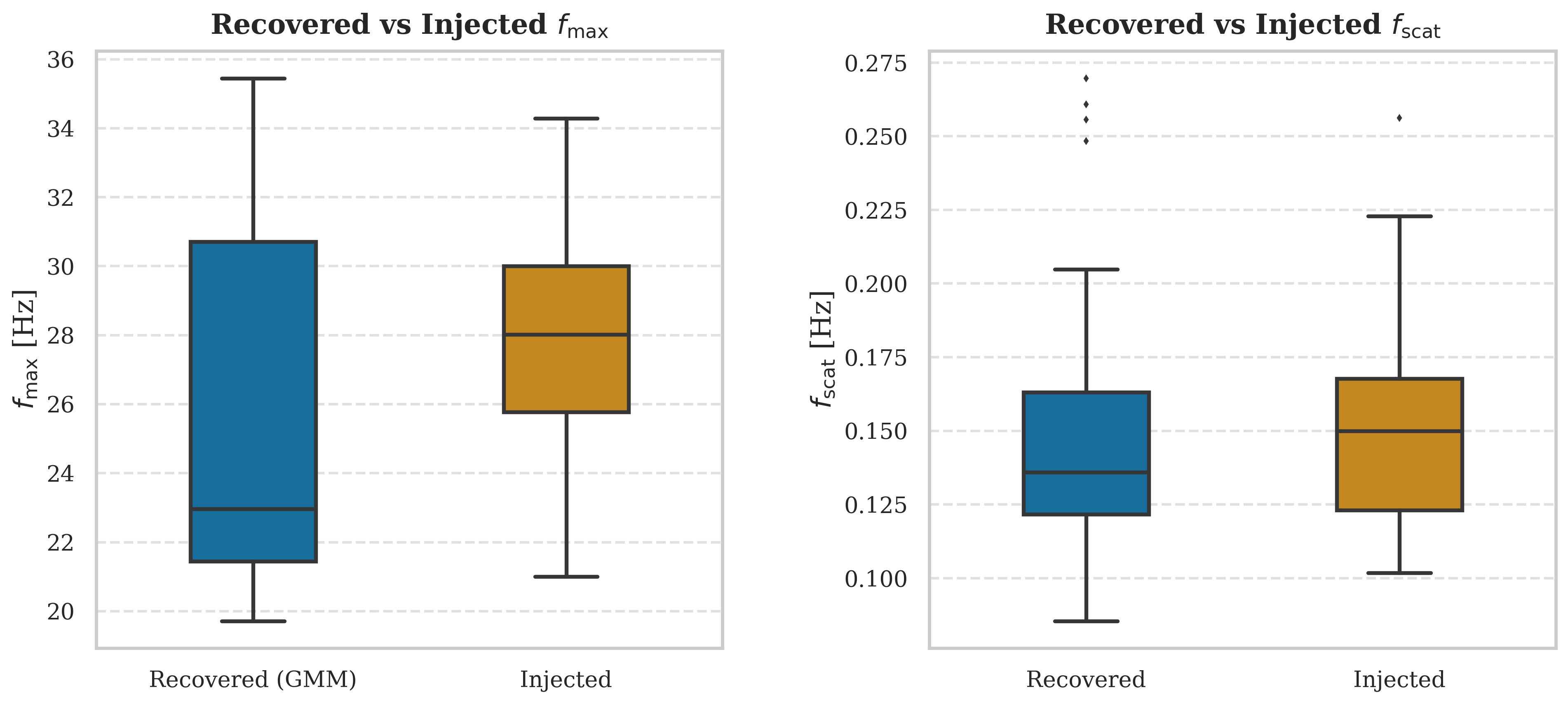}
\caption{Recovered properties of scattering noise (using ArchGEM) vs. injected scattering noise (with GlitchPop, which also includes IMBH signal). Left plot refers to the maximum scattering frequency and the plot on right shows the frequency of scattering. 
    }
\label{fig:Arch_sim}
\end{figure*}

In this simulated dataset, each sample contains one scattered light episode with two arches in the 10--60~Hz frequency band, along with a simulated GW injection, yielding a diverse range of frequency-time trajectories. The results of this validation are summarized in Figure~\ref{fig:Arch_sim}. We find that \textsc{ArchGEM} successfully recovers scattering frequencies $f_{\mathrm{scat}}$ in the range 0.1--0.2~Hz, consistent with the injected glitches. This agreement demonstrates that the algorithm reliably captures the temporal periodicity of successive arches—the defining signature of scattered-light motion. The recovered maximum frequencies $f_{\max}$ are typically higher than those in the injected set (median $\sim$38~Hz versus 28~Hz), reflecting that ArchGEM measures the upper envelope of broadband energy distributions in the spectrogram, whereas the injected $f_{\max}$ values represent single narrowband modulation frequencies. Real scattered-light arches often exhibit blended harmonic content and multi-path interference, leading to higher apparent $f_{\max}$ in the recovered distribution. Because $v_{\mathrm{surf}}$ is directly proportional to $f_{\max}$, this systematic difference naturally propagates to higher inferred surface velocities in the ArchGEM results. Furthermore, as shown in \citep{McGowan2025archgem}, ArchGEM has been extensively tested on numerous scattered-light cases throughout O4 and has been shown to reliably recover low-frequency scattered-light glitches. The validation study presented here on high-mass injections further demonstrates ArchGEM's effectiveness even in the presence of signals occupying similar low-frequency bands. \\

For both simulated and recovered datasets, the physical quantities $|x_{\mathrm{surf}}|$ and $|v_{\mathrm{surf}}|$ are calculated using standard relations, with $f_{\mathrm{scat}}$ representing the scattering recurrence frequency and $f_{\max}$ the maximum observed modulation frequency. In the injected data, these parameters are inferred from the known input period specified in \textsc{GlitchPop}; while the injection pipeline does not explicitly compute $x_{\mathrm{surf}}$ or $v_{\mathrm{surf}}$, these quantities can be derived analytically from the injected modulation frequency for direct comparison to ArchGEM’s recovered values. Importantly, ArchGEM itself operates entirely independently of the injected parameters. It only uses the time-frequency morphology present in the generated Q-transforms, and its GMM-based selection ensures that only statistically significant peaks contribute to the recovered distributions. The close agreement between the recovered and injected distributions confirms that ArchGEM identifies the correct scattering periodicity and does not respond to the coincident GW signal, demonstrating its selectivity and robustness in mixed astrophysical and instrumental data. From these recovered parameters, \textsc{ArchGEM} infers physical surface displacements $|x_{\mathrm{surf}}|$ of roughly 75--200~$\mu$m and corresponding average surface velocities $|v_{\mathrm{surf}}|$ of 15--22~$\mu$m~s$^{-1}$. These amplitudes are consistent with expected motion of optical baffles, suspended optics, and other scattering surfaces under typical environmental excitation at LIGO sites. The agreement between simulated and recovered quantities confirms that \textsc{ArchGEM} correctly maps spectro-temporal features into physically meaningful motion parameters. Each 9-second strain requires approximately 10~minutes to process on the LDAS cluster (\texttt{ldas-pcdev5.ligo.caltech.edu}), utilizing 512~GB of memory and 40~CPU cores. The majority of runtime is spent computing the high-resolution Q-transform and subsequent Gaussian mixture modeling of the energy clusters, followed by multi-channel spectrogram generation and metric extraction. Despite this computational expense, \textsc{ArchGEM}'s performance demonstrates that automated, statistically robust scattered-light analysis is feasible for full-run datasets when parallelized across compute nodes. \\


Next, we evaluated AWaRe's waveform reconstruction performance on the simulated IMBH dataset. To do so, we first tested the network's ability to produce calibrated uncertainty estimates alongside accurate reconstructions. For each injection, the model outputs a mean waveform ($\mu$) and a corresponding per-sample standard deviation ($\sigma$), representing its predictive uncertainty. The calibration test follows a pointwise coverage analysis for Gaussian predictive marginals. For each time sample in the reconstructed waveform, the predicted $(\mu, \sigma)$ pairs are compared with the true waveform value ($y_{\mathrm{true}}$). For a given nominal confidence level $c$ (e.g., 10\%, 50\%, 90\%), we compute the corresponding Gaussian quantile $z_c = \Phi^{-1}[(1 + c/100)/2]$, where $\Phi^{-1}$ is the inverse cumulative distribution function. A point is considered ``covered” if $y_{\mathrm{true}}$ lies within the interval $\mu \pm z_c\sigma$. The empirical coverage is then obtained as the fraction of all samples that satisfy this condition, averaged over all injections in the test set. Plotting the empirical coverage against the nominal coverage provides a quantitative measure of calibration, with a perfectly calibrated model lying along the 1:1 diagonal. \\

\begin{figure*}[ht!]
\gridline{\fig{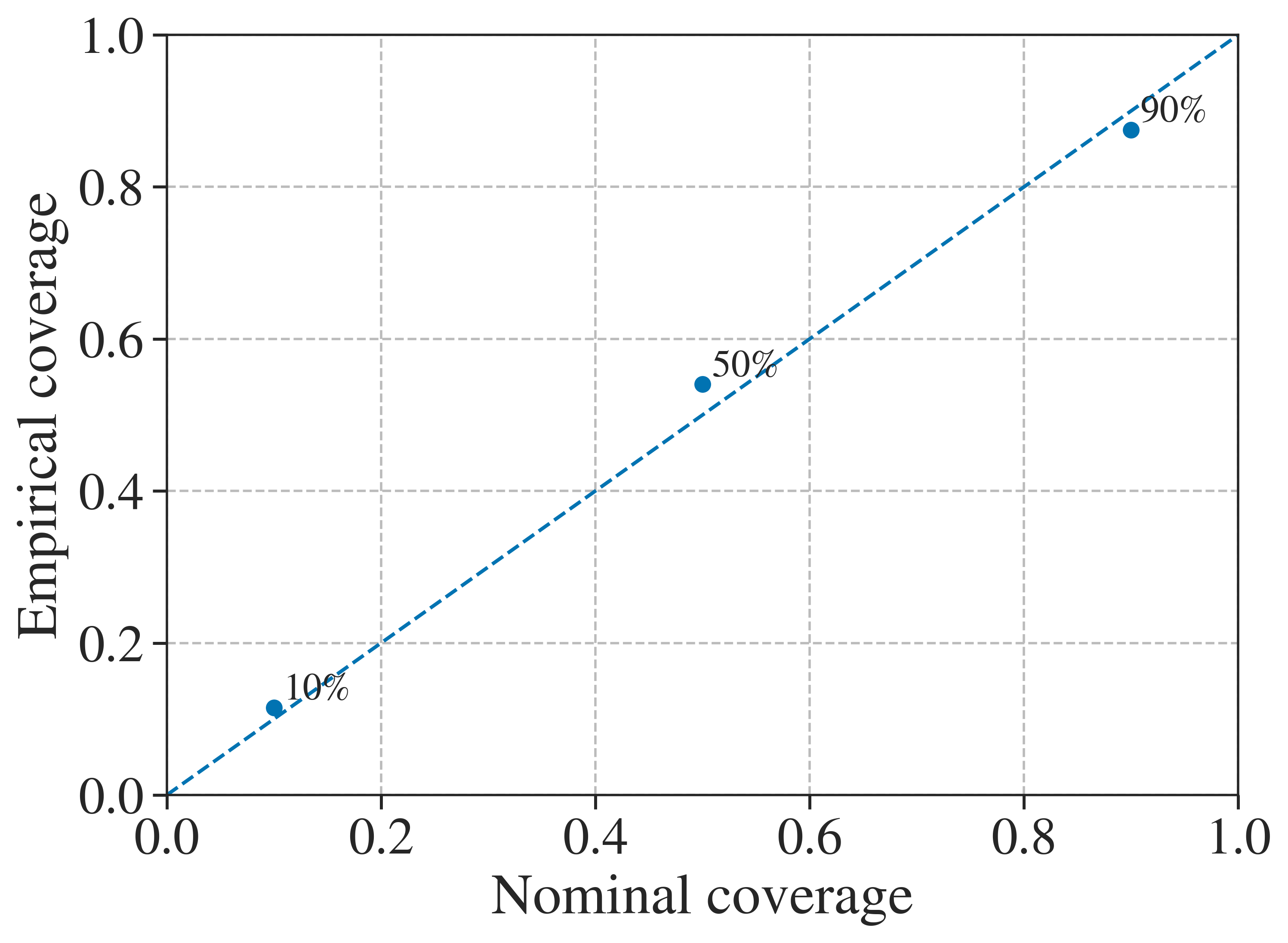}{0.52\textwidth}{(a)}
\fig{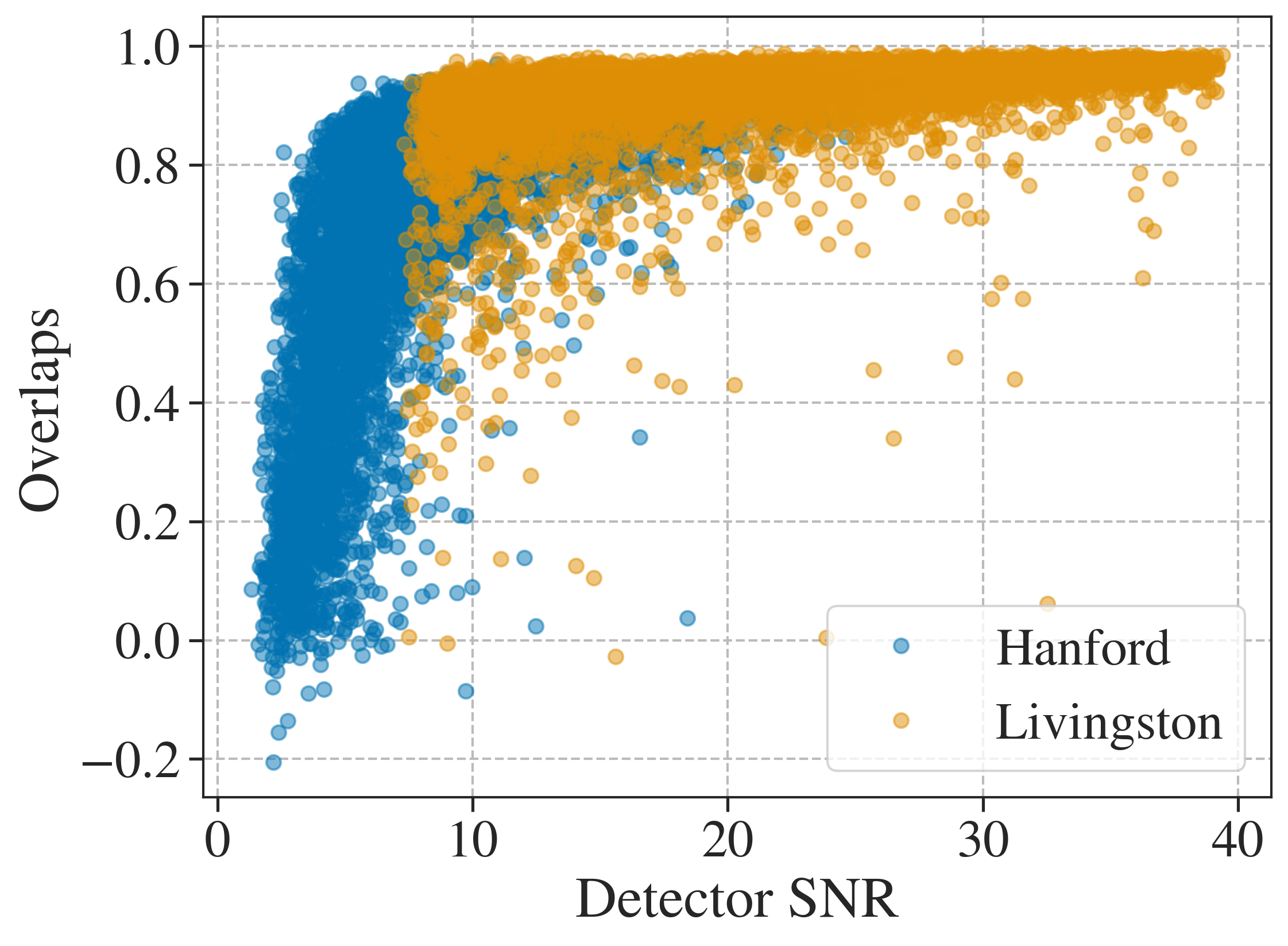}{0.48
  \textwidth}{(b)}
}
\caption{\label{fig:AWaRe_IMBH_test} (a) Empirical versus nominal coverage for AWaRe predictive uncertainties, evaluated on the simulated IMBH dataset. The points correspond to the central 10\%, 50\%, and 90\% confidence intervals of the Gaussian predictive distributions. (b) Overlap between reconstructed and true waveforms as a function of the single detector SNR for the Hanford and Livingston detectors.}
\end{figure*}

Fig.~\ref{fig:AWaRe_IMBH_test} (a) shows the resulting empirical-versus-nominal coverage curve. For this test, we injected the same waveforms as in the GW-Whisper test in public O4 data, while the model is trained on injections in O3 noise. The results lie close to the 1:1 line, demonstrating that AWaRe’s uncertainty estimates are well calibrated for IMBH-scale signals. Minor deviations at higher confidence levels indicate a slight underestimation of variance in high-SNR regions, consistent with conservative uncertainty predictions. Overall, this test confirms that AWaRe achieves statistically reliable reconstruction with meaningful uncertainty quantification across the simulated IMBH parameter space.

To further evaluate waveform reconstruction performance, we examined how the overlap between the reconstructed and true waveforms varies with the single detector SNR (Fig.~\ref{fig:AWaRe_IMBH_test} (b)). For each simulated IMBH injection, the overlap was computed independently for the Hanford and Livingston detectors. Figure 8 shows the resulting distribution of overlaps as a function of detector SNR. Reconstruction fidelity increases with SNR, with overlaps exceeding 0.9 for most injections with SNR $\gtrsim 10$. A small subset of cases with SNR $>10$ exhibit overlaps between 0.0 and 0.1. These correspond to instances where the injected signal coincides with a glitch, leading the model to partially reconstruct the glitch waveform. This behavior is expected, since IMBH signals can exhibit morphologies similar to many glitch classes like blips and tomte. At higher SNRs, the overlaps saturate near unity, indicating that AWaRe accurately captures both the phase and amplitude evolution of IMBH-scale signals across detectors.

\section{Conclusions}  
We have presented the first ML follow-up of the BBH merger GW231123. Our integrated pipeline, comprising \textsc{GW-Whisper} for low-latency segment classification, \textsc{ArchGEM} for scattered-light glitch diagnostics, and \textsc{AWaRe} for probabilistic waveform reconstruction, addresses simultaneously the dual challenges that defined this event: strong waveform-model systematics and nearby non-Gaussian noise transients. \textsc{GW-Whisper} autonomously flags the merger without human vetting. \textsc{ArchGEM} extracts the physical parameters of the scattered-light arches, providing feedback for detector commissioning. \textsc{AWaRe} produces reconstructions that agree more closely with template-free cWB and BayesWave results than with template-based Bilby posteriors. Injection studies covering seven common glitch morphologies and an extended mass range of $100$–$1000\,M_\odot$ show that \textsc{AWaRe} retains high fidelity across the lower-mass IMBH regime, and quantifies the onset of accuracy loss for the heaviest systems. We extended the validation study to test the generalizability of all three models under realistic detector conditions. The fine-tuned \textsc{GW-Whisper} model, originally trained on the MLGWSC-1 dataset, demonstrated strong performance on IMBH-scale signals (AUROC = 0.977, AUPRC = 0.997) and maintained a $\mathrm{FAR} < 100$ per month for most injections. The \textsc{AWaRe} reconstruction overlaps on the same dataset exceeded 0.9 for most signals with SNR $\gtrsim 10$. Complementing this, a dedicated scattered light injection study demonstrated that \textsc{ArchGEM} reliably recovers the injected scattering recurrence and maximum frequencies ($f_{\mathrm{scat}}$ and $f_{\max}$) within the expected statistical range. 
Together, these results demonstrate that the combined \textsc{GW-Whisper–ArchGEM–AWaRe} framework remains robust and statistically reliable, establishing a foundation for scalable, real-time inference in the IMBH mass range.  \\

\section{Code and Data Availability}

The codes for training and reproducing the analysis are available on this GitHub \href{https://github.com/chayanchatterjee/GW231123-ML-DetChar}{repo}. The GW231123 data and the background noise from O3 and O4 were obtained from publicly available data in GWOSC \citep{GWOSC}. The injection sets for the validation studies reported in this paper are available on Zenodo (\href{https://doi.org/10.5281/zenodo.17459199}{https://doi.org/10.5281/zenodo.17459199}).

\begin{acknowledgments}
The authors would like to thank Tabata Ferreira for helpful comments and suggestions. This research was undertaken with the support of compute grant and resources, particularly the DGX A100 AI Computing Server, offered by the Vanderbilt Data Science Institute (DSI) located at Vanderbilt University, USA. This research used data obtained from the Gravitational Wave Open Science Center (https://www.gw-openscience.org), a service of LIGO Laboratory, the LIGO Scientific Collaboration and the Virgo Collaboration. LIGO is funded by the U.S. National Science Foundation. Virgo is funded by the French Centre National de Recherche Scientifique (CNRS), the Italian Istituto Nazionale della Fisica Nucleare (INFN) and the Dutch Nikhef, with contributions by Polish and Hungarian institutes. This material is based upon work supported by NSF's LIGO Laboratory which is a major facility fully funded by the National Science Foundation.
\end{acknowledgments}

%




\bibliography{sample701}{}
\bibliographystyle{aasjournalv7}



\end{document}